# Advances in Data Combination, Analysis and Collection for System Reliability Assessment

**Alyson G. Wilson, Todd L. Graves, Michael S. Hamada and C. Shane Reese**

*Abstract.* The systems that statisticians are asked to assess, such as nuclear weapons, infrastructure networks, supercomputer codes and munitions, have become increasingly complex. It is often costly to conduct full system tests. As such, we present a review of methodology that has been proposed for addressing system reliability with limited full system testing. The first approaches presented in this paper are concerned with the combination of multiple sources of information to assess the reliability of a single component. The second general set of methodology addresses the combination of multiple levels of data to determine system reliability. We then present developments for complex systems beyond traditional series/parallel representations through the use of Bayesian networks and flowgraph models. We also include methodological contributions to resource allocation considerations for system reliability assessment. We illustrate each method with applications primarily encountered at Los Alamos National Laboratory.

## 1. INTRODUCTION

By definition, *reliability* is the probability a *system* will perform its intended function for at least a given period of time when operated under some specified conditions. The systems that we are asked to assess are becoming increasingly complex, including, for example, nuclear weapons, infrastructure networks, supercomputer codes and munitions. In many instances it is not possible to mount vast numbers of full system tests, and frequently none is available (Bennett, Booker, Keller-McNulty and Singpurwalla, 2003). Systems reliability methodology is faced with the challenge of developing models for these complex systems and integrating multiple, sometimes indirect, sources of information to perform estimation, make inferences and answer questions about the allocation of additional testing resources.

*Alyson Wilson, Todd Graves and Michael Hamada are Technical Staff Members, Los Alamos National Laboratory, Los Alamos, New Mexico 87545, USA e-mail:*
*agw@lanl.gov; tgraves@lanl.gov; hamada@lanl.gov.*
*C. Shane Reese is Associate Professor, Department of Statistics, Brigham Young University, Provo, Utah 84602, USA e-mail: reese@stat.byu.edu.*

This paper focuses on four methodological issues that arise from complex systems reliability problems. In Section 2, we address methods for integrating multiple data sources to assess the reliability of a single component. The data may come from many sources, including experimental test results, computer simulations and expert opinion. In Section 3, we consider methods for assessing systems reliability when the data are available at multiple levels (e.g., both system and component). Again, there may be multiple sources of data at each component or at the system itself. In Section 4, we discuss Bayesian networks and flowgraph models, which are richer representations that are able to model more systems than fault trees or reliability block diagrams can. In Section 5 we consider the resource allocation problem for systems. Section 6 summarizes our view of some of the current research challenges in systems reliability assessment.

The analyses presented here follow hierarchical Bayesian approaches and focus on estimating the reliability $R(t)$, in most cases as a function of time. We will write $R(t|\Theta)$ to denote reliability given unknown parameters $\Theta$, and after obtaining a posterior distribution $\pi(\Theta|D)$ for $\Theta$ based on data $D$, estimates of $R(t)$ can be obtained from, for example, the posterior mean $\int R(t|\Theta)\pi(\Theta|D)\,d\Theta$. In each example we use these meanings for $R, \pi, \Theta$ and $D$.

## 2. INTEGRATING MULTIPLE DATA SOURCES TO ASSESS COMPONENT RELIABILITY

In this section, we consider the assessment of component reliability when multiple data sources are available. Ideally, we would like a large set of pass/fail tests or failure time observations to estimate the reliability of a component. We are often in situations where this is not the case, but we are able to supplement our data with other information sources. In this section, we consider specifically degradation data, surrogate data and a biased sample of pass/fail data.

### 2.1 Degradation and Failure Time Data

An important practical example is the case where failure time data are augmented with degradation data. Suppose that we are interested in the lifetime distribution of a component. In the past we have observed $n_1$ failures at times $T_j$ for $j = 1, \ldots, n_1$. A further $n_2$ components are still functioning and their ages are $A_j$ for $j = n_1 + 1, \ldots, n_1 + n_2$. Finally, $n_3$ components were destructively tested and these tests yielded the continuous measurements $Y_j$ at ages $t_j$ for $j = n_1 + n_2 + 1, \ldots, n_1 + n_2 + n_3$. The $Y_j$ tend to decrease with age and it is thought that this decrease is closely related to the eventual failure of the components.

We seek to analyze these data simultaneously using a hierarchical Bayesian approach by first assuming that the degradation process satisfies

$$Y_j \sim \text{Normal}(\alpha - \beta_j^{-1} t_j, \sigma_y^2).$$

This assumption implies that components are identical at birth, although measurement error is present even when testing new units. Differences in components arise later as each is allowed to degrade at its own rate $\beta_j^{-1}$, and we assume that $\log \beta_j \sim \text{Normal}(\mu, \sigma_b^2)$. We estimate both $\mu$ and $\sigma_b$; $\mu$ has a normal prior distribution and $\sigma_b$ has a gamma prior distribution. The measurement error standard deviation $\sigma_y$ is also given a gamma prior distribution. To relate this degradation process to the failure times, assume that a critical lower level $L$ exists and that

$$T_j = \inf\{t \geq 0 : \alpha - \beta_j^{-1} t \leq L\} = (\alpha - L)\beta_j,$$

so that $\log T_j \sim \text{Normal}(\mu + \log(\alpha - L), \sigma_b^2)$. In this problem the reliability is defined to be the survivor function of a generic lifetime $T$, $P\{T > t\}$. The level $L$ can be given a prior distribution and estimated; in most cases the value of the degradation process that is required for successful performance will be approximately known, so that this prior distribution will be informative. We assume that $L/\alpha \sim \text{Beta}(a, b)$. The lognormal distribution for $T_j$ defines the likelihood for both the censored lifetimes and the observed lifetimes. This yields the unnormalized posterior distribution

$$
\begin{aligned}
&\pi(\Theta|D)\\
&= \pi(\alpha, \beta, \sigma_b, \mu, \sigma_y, \mathbf{L}|\mathbf{T}, \mathbf{A}, \mathbf{y})\\
&\propto \phi\left(\frac{\alpha - m_\alpha}{s_\alpha}\right)\phi\left(\frac{\mu - s_\mu}{s_\mu}\right)\sigma_y^{a_{\sigma_y}-1}\exp(-r_{\sigma_y}\sigma_y)\\
&\quad\cdot \sigma_b^{a_{\sigma_b}-1}\exp(-r_{\sigma_b}\sigma_b)\\
&\quad\cdot \alpha^{-1}\left(\frac{L}{\alpha}\right)^{a-1}\left(\frac{\alpha - L}{\alpha}\right)^{b-1}\\
&\quad\cdot \prod_{j=1}^{n_1}((\sigma_b T_j)^{-1}\phi[\{\log T_j - \mu\\
&\hspace{4cm} - \log(\alpha - L)\}/\sigma_b])\\
&\quad\cdot \prod_{j=n_1+1}^{n_1+n_2}(1 - \Phi[\sigma_b^{-1}\{\log A_j - \mu\\
&\hspace{4cm} - \log(\alpha - L)\}])
\end{aligned}
$$

(1)

$$\cdot \prod_{j=n_1+n_2+1}^{n_1+n_2+n_3} [(\sigma_b \beta_j)^{-1} \phi\{(\log \beta_j - \mu)/\sigma_b\}\sigma_y^{-1}$$

$$\cdot \phi(\{y_j - \alpha - \beta_j^{-1}t_j\}/\sigma_y)],$$

where $\phi$ and $\Phi$ denote the standard normal density and distribution functions, respectively, and where $m_\alpha, s_\alpha, m_\mu, s_\mu, s_{\sigma_y}, r_{\sigma_y}, s_{\sigma_b}, r_{\sigma_b}, a$ and $b$ denote fixed quantities that define prior distributions for $\alpha$ and other parameters. Samples from this unnormalized posterior distribution can be drawn using a variable-at-a-time random walk Metropolis algorithm.

As an example, consider a simulated population of items at time 20 years after fabrication. We have observed four failures, all in the last two years, and 76 items have survived to this point. We also have one degradation data point per year up to year 20. The data were simulated under $\alpha = 100$, $L = 20$ and $\mu = \log(0.35) = -1.05$; this implies that the degradation curve will cross level $L$ at age $0.35(100 - 20) = 28$ years. Other parameters of the simulation include $\sigma_b = 0.2$ and $\sigma_y = 5$. In our prior distributions, we used $\alpha \sim \text{Gamma}(4, 1/30)$ (with mean 120), $\sigma_y \sim \text{Gamma}(4, 1/2.5)$, $\sigma_b \sim \text{Gamma}(4, 5)$, $\mu \sim \text{Normal}(0, 1)$, and $L/\alpha \sim \text{Uniform}(0, 1)$. The results are shown in Figure 1. The solid curve is the true reliability (survivor function) $R(t)$, the dashed curve is $\int \Phi(\frac{\mu + \log(\alpha - L) - \log t}{\sigma_b})\pi(\Theta|D)$, the posterior mean of the survivor function, and the dotted curves are the 5th and 95th percentiles of the posterior distribution. There is substantial uncertainty in the reliability just a few years into the future, but this is considerably better than could be obtained using the (mostly censored) failure times alone. Posterior estimates (and 90% posterior probability intervals) for the parameters are 99.2 (92.9, 105.1) for $\alpha$, 17.6 (2.3, 34.6) for $L$, $-1.00$ ($-1.21$, $-0.76$) for $\mu$, 6.57 (3.8, 10.3) for $\sigma_y$ and 0.24 (0.14, 0.35) for $\sigma_b$. This approach has the advantage that the threshold parameter $L$ does not need to be known with certainty and can be estimated; doing so can provide a diagnostic for the value historically assumed for $L$. The approach can also benefit from strong prior information about $L$, which might come from physical or engineering knowledge used to define the requirements for the component.

## 2.2 Bernoulli and Quality Assurance Data

Anderson-Cook et al. (2005) applied ideas from medical statistics to combine pass/fail data with component quality assurance data to get more precise reliability estimates. Anderson-Cook et al. (2005) actually worked in a system context but here we discuss the single component variant of the problem; see Section 3.2 for the system extension. A component undergoes destructive pass/fail testing at various ages. Suppose that age is the only covariate of interest, although the model is general enough to allow multiple covariates. Further suppose that the component can also be tested destructively for adherence to up to $J$ published specifications. We assume that each such test related to the $j$th specification ($j = 1, \ldots, J$) yields (possibly after transformation) a normally distributed measurement with mean $\alpha_j + \delta_j t$ and variance $\gamma_j^2$ if the test is conducted at age $t$. It is thought that these specification measurements are related to the component's performance in a pass/fail test, and we assume that the measurements have been transformed so that large values of the measurement are thought to be good. We now invoke an assumption to relate the two types of data. This assumption is inspired by the concept of surrogate variables in medical studies (Prentice, 1989; Pepe, 1992). Suppose that it were possible to obtain a system test $Y$ on the same unit where we obtained a full set of specification measurements $Z_1, \ldots, Z_J$. Then we assume

$$\Pr\{Y = 1|Z, t\} = \prod_{j=1}^{J} \Phi\left(\frac{Z_j - \theta_j}{\sigma_j}\right),$$

independently of $t$. In this model, each of the $J$ quantities represented in specifications is independently capable of causing failure, and it is not possible, for example, for two quantities with somewhat low values to collaboratively cause failure. If the latter behavior is desired, it is possible to replace the product with a multivariate normal integral. Here $\theta_j$ and $\sigma_j$ are unknown, given prior distributions, and estimated. Their prior distributions can be informative if the published specifications are thought to be highly relevant to reliability. The key result, since it is impossible to observe the $Z_j$ for a component that undergoes pass/fail testing, is that the $Z_j$ can be integrated out, so that

$$(2) \quad \begin{aligned} R(t|\Theta) &= \Pr\{Y = 1|t, \Theta\} \\ &= \prod_{j=1}^{J} \Phi\left(\frac{\alpha_j + \delta_j t - \theta_j}{\sqrt{\gamma_j^2 + \sigma_j^2}}\right). \end{aligned}$$

Terms like this can be multiplied by normal density terms that reflect the specification measurements so as to combine the two sources of data. Assuming that the data consist of system tests $Y_1, \ldots, Y_m$ taken at ages $t_1, \ldots, t_m$ and specification measurements $Z_1, \ldots, Z_n$ taken at ages $\tau_1, \ldots, \tau_n$, where measurement $Z_i$ corresponds to the $k_i$th specification, the likelihood function is

$$L(\alpha, \delta, \gamma, \sigma, \theta | \mathbf{Y}, \mathbf{Z})$$
$$= \prod_{i=1}^{m} R(t_i | \Theta)^{Y_i} \{1 - R(t_i | \Theta)\}^{1-Y_i}$$
$$\cdot \prod_{j=1}^{n} \sigma_{k_j}^{-1} \phi\left(\frac{Z_j - \alpha_{k_j} - \delta_{k_j} \tau_j}{\sigma_{k_j}}\right).$$

### 2.3 Biased and Unbiased Samples

Graves et al. (2006) discussed a challenging problem whose solution could be applied in a reliability context because it involves the estimation of prevalence of a feature in a stratified population. A population of items was manufactured in lots, and it was of interest to estimate the fraction of items in each lot with a certain feature. There was reason to believe that feature prevalence had a nonzero, but imperfect, relationship with lot membership, so the authors assumed that if the $j$th lot was of size $N_j$, the number of features $K_j$ in the lot had a Binomial$(N_j, p_j)$ distribution, where the $p_j$ had a hierarchical prior $p_j \sim \text{Beta}(a, b)$, with $a$ and $b$ given prior distributions. Some of the lots were inspected using random (hypergeometric) sampling: a sample of size $n_j^r$ was taken from the $j$th lot for inspection and $y_j^r$ features were found. These data alone can be analyzed using a Markov chain Monte Carlo (MCMC) algorithm to obtain samples from the joint distribution of $(a, b, \mathbf{p}, \mathbf{K})$. However, some other feature data were available from items selected using nonrandom sampling (a "convenience sample"); the selection process may or may not have been independent of feature presence. To combine these two sources of data, one needs to model this nonrandom sampling, and Graves et al. (2006) used the extended hypergeometric distribution. In fact, the convenience samples were taken before the random samples. Denote by $n_j^c$ and $y_j^c$ the sample size and number of features found from the $j$th lot in the convenience sample. Then the extended hypergeometric model is

$$P(y_i^c = y)$$
$$= \frac{\binom{n_i^c}{y} \binom{N_i - n_i^c}{K_i - y} \theta^y}{\sum_{j=\max(0, n_i^c - N_i + K_i)}^{\min(n_i^c, K_i)} \binom{n_i^c}{j} \binom{N_i - n_i^c}{K_i - j} \theta^j}.$$

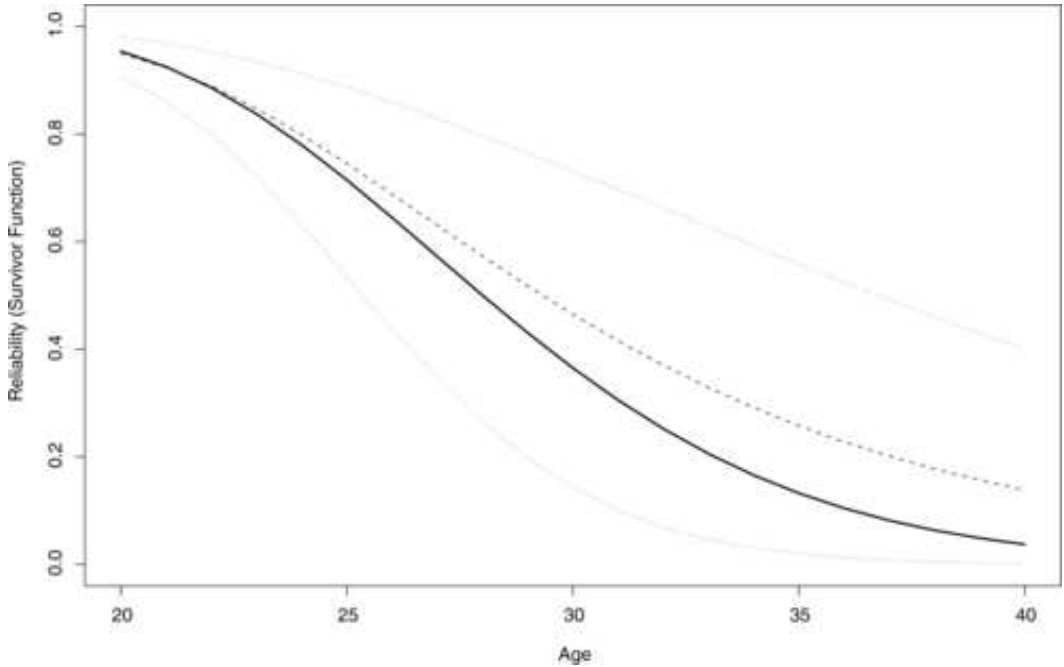

FIG. 1.   *Reliability estimates with uncertainty bands for the degradation and failure time data integration example. The solid curve is the true reliability function, the dashed curve is the posterior mean and the dotted curves are the 5th and 95th percentiles of the posterior distribution.*

When the unknown biasing parameter $\theta = 1$, this is the hypergeometric model; for $\theta > 1$, items with the feature are more likely to be sampled, and so forth. Graves et al. (2006) assumed that the amount of biasing is constant in each lot ($\theta$ does not depend on the lot), put a lognormal prior distribution on $\theta$ and estimated the amount of biasing. Their data set turned out to be inconclusive about the direction of the bias. The likelihood for the randomly sampled data is

$$y_i^r \sim \text{Hypergeometric}(K_i^r, N_i^r - K_i^r, n_i^r),$$

which is to say

$$P(y_i^r = y) = \frac{\binom{n_i^r}{y}\binom{N_i^r - n_i^r}{K_i^r - y}}{\binom{N_i^r}{K_i^r}},$$

where $N_i^r = N_i - n_i^c$ and $K_i^r = K_i - y_i^c$. Graves et al. (2006) sampled from the resulting posterior distribution of $(\mathbf{p}, \mathbf{K}, a, b, \theta)$ using YADAS. Integrating the convenience samples with the randomly sampled data enabled a more precise estimate of the quantity of interest—the prevalence of features among the unsampled items, $f(\mathbf{K}) = \sum_i (K_i - y_i^c - y_i^r)/\sum_i (N_i - n_i^c - n_i^r)$—without making unwarranted assumptions such as the prevalence of features being the same in each lot or the convenience sampling being done independently of feature presence. In a simplified case where items that lack the feature have reliability 1 and items with the feature have reliability 0, the posterior mean reliability is the integral of $f(\mathbf{K})$ with respect to the posterior distribution of $\mathbf{K}$.

Further study is required before one can recommend using a more informative prior for the amount of bias $\theta$. It is difficult to relate the parameter to knowledge about the sampling process in a quantitatively precise manner. If the biasing mechanism is better understood, that mechanism should be explicitly included in the model rather than the approach given here.

## 3. ASSESSING SYSTEM RELIABILITY WITH MULTILEVEL DATA

In Section 2, we discussed combining multiple data sources to assess a single component. In this section, we consider combining multiple sources of data in a system reliability assessment. In particular, we consider situations where we have data about both components and combinations of components—for example, about the entire system. Hamada et al. (2004) developed models for the case of a fault tree with binary data at basic, intermediate and top events. Here we give examples of combining failure time data, failure count data, Bernoulli data and degradation data.

### 3.1 Logistic Regression, Weibull Lifetimes and Degradation

As an example of integrating multilevel reliability data, we work with a variant of an analysis discussed in Graves and Hamada (2005). The system consists of three components combined in series, and all three components may see degrading performance with age. For component 1, we have binary test data at various ages and we assume a logistic regression relationship for the success probability as a function of age. If $X_1$ denotes a generic component 1 of age $t$ (centered) and $X_1 = 1$ denotes component success,

$$\text{logit } \Pr\{X_1 = 1\} = \theta_0 + \theta_1 t.$$

We assume independent normal priors for $\theta_0$ and $\theta_1$, and in our simulated data, we have 25 tests each at ages 0, 2, 4, 6, 8, 10, 15 and 20, with one failure at age 4, two at age 15 and six at age 20.

Component 2 is assumed to have a Weibull lifetime distribution with

$$\Pr\{T_2 > t\} = \exp(-\lambda_0 t^{\lambda_1}),$$

where $T_2$ denotes a generic lifetime for component 2. Component 2 is said to work properly in a test if its life has not yet ended at the time it is tested. We observe eight uncensored lifetimes ranging from 14.1 years to 33.5 years, with 13 lifetimes right-censored at 20 years and four right-censored at 40 years.

Our data for component 3 mirrors the analysis in Section 2.1: we have ten total pieces of degradation data taken every two years [these data are normal with mean $\alpha + \beta_j^{-1} t_j$ and variance $\sigma_{y}^2$, with $\log \beta_j \sim \text{Normal}(\mu, \sigma_b^2)$] and 80 lifetimes, all but two of them censored at 20 years. [The logs of these data are normal with mean $\mu + \log(\alpha - D)$ and variance $\sigma_b^2$.] This time, we assume that $D = 20$ is known with certainty. Finally, we also have binomial system test data (15 tests each at ages 0, 5, 10, 15 and 20, with one failure at age zero and three at age 20). Since this is a series system, the probability of system success for a system of age $t$ is then

$$R(t|\Theta) = \text{logit}^{-1}(\theta_0 + \theta_1 t) \cdot \exp(-\lambda_0 t^{\lambda_1})$$
$$\cdot \{1 - \Phi(\{\log t - \mu - \log(\alpha - D)\}/\sigma_b)\}.$$

The component data sets can be analyzed together with the system data by multiplying all the likelihood functions with the prior distributions for all the unknown parameters. Again, samples from the posterior distribution can be drawn using a variable-at-a-time random walk Metropolis algorithm, and setting up the problem is straightforward in YADAS (Graves, 2003). YADAS can handle much larger systems (e.g., Johnson, Graves, Hamada and Reese, 2003), for the case of pass/fail data with no aging at all levels). The user can specify the system structure in a file and component data can take many forms, assuming only that the user can express the success probability at each component as a function of unknown parameters. Figure 2 displays the results of the analysis. For each component and for the full system, we display the mean and 5th and 95th percentiles of the posterior distribution of reliability as a function of age. Component 2 dominates the unreliability at early ages, while the other two components are bigger concerns at later ages.

### 3.2 Combining Partially Informative System Tests with Component Tests

Anderson-Cook et al. (2005) analyzed data from a system in which the system pass/fail testing data provide incomplete information about which component(s) was responsible for a failure. In particular, for the $i$th test, the data consist of a set $C_1(i)$ of components known to have worked, a second set of components $C_2(i)$ known to have failed, and a third set of components $C_3(i)$, where it is known that at least one component in that set failed. Anderson-Cook et al. (2005) did this in the context of combining these system tests with component specification testing data (see Section 2.2). In a multiple component context, denote by $p_{ik}$ the probability in (2) that component $k$ works properly in test $i$. Then the probability of observing data $(C_1(i), C_2(i), C_3(i))$ given these component success probabilities is

$$\left\{ \prod_{k \in C_1(i)} p_{ik} \right\} \left\{ \prod_{k \in C_2(i)} (1 - p_{ik}) \right\} \left\{ 1 - \prod_{k \in C_3(i)} p_{ik} \right\},$$

where the third product is understood to equal 0 if it is empty (the other products are 1 if empty). Results obtained by Anderson-Cook et al. (2005) for a two-component series system are shown in Figure 3. Denoting by $R_i(t|\Theta)$ the reliability of component $i$ given in expression (2), the posterior mean system reliability is $\int R_1(t|\Theta) R_2(t|\Theta) \pi(\Theta|D) \, d\Theta$. Since the data are proprietary, both axes (time and reliability) have been rescaled to $[0, 1]$. The black curves show the integration of the two types of data (posterior means, 5th and 95th percentiles of the posterior distribution). The solid and dashed curves show the previous methodology used by the system engineers: logistic regression using full system data only. The component test data are in this case available for older components, which greatly tightens the uncertainty bounds for older systems (dotted lines). (This analysis depicts a small subsystem of the full system, and none of the components in the small subsystem appears to age significantly.)

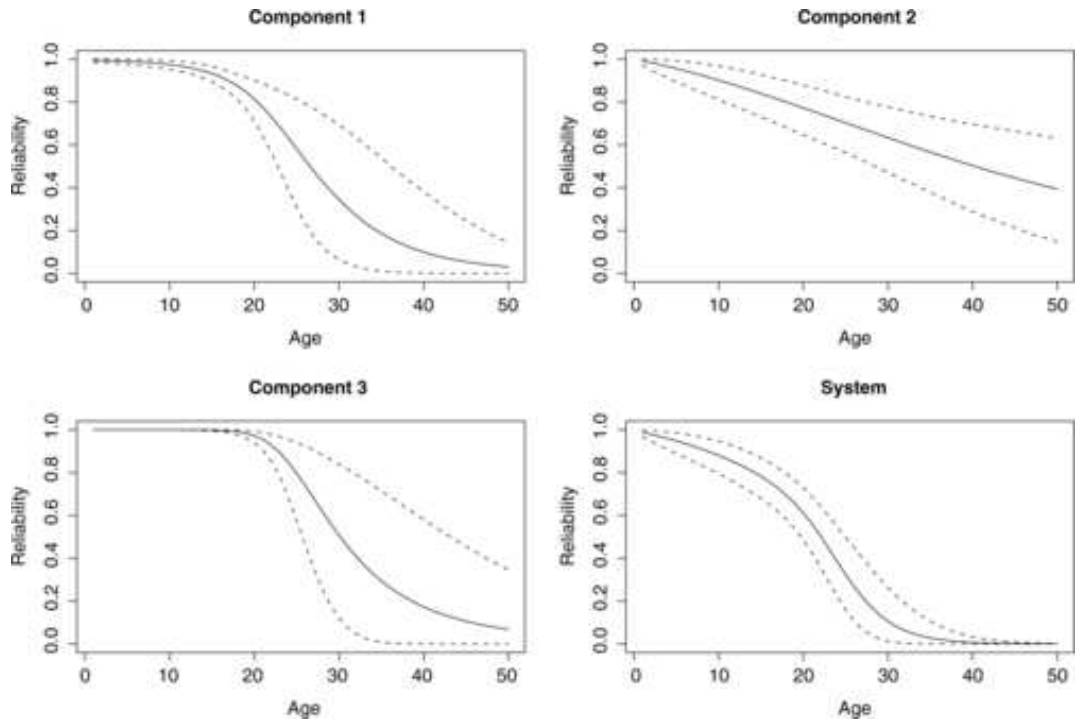

Fig. 2. *Reliability estimates and uncertainty intervals for the three-component system. Upper left: Component 1, which has logistic regression data. Upper right: Component 2, with Weibull failure time data. Lower left: Component 3, with both degradation data and lognormal failure time data. Lower right: The full series system with all four data sets.*

This is a form of "autopsy data." Meilijson (1994) used the expectation-maximization algorithm to obtain maximum likelihood estimates for failure time distribution parameters from the failure time of the system and the set of components that failed by that time. Gåsemyr and Natvig (2001) worked with lifetime data, where the set of failed components is identified when the system fails and some components are monitored either at all times or from certain time points onward (if a component fails while being monitored, its failure time is observed exactly). They also observed systems that did not fail before a censoring time. They derived expressions for the likelihood function under general system structures, including the case of dependent failures, and identified conjugate prior distributions in the case that failure times follow generalized gamma distributions.

### 3.3 Nonhomogeneous Poisson Process

Highly clustered modern supercomputers are examples of systems composed of many similar systems in series. Ryan and Reese (2005) presented a model for the reliability of a Los Alamos National Laboratory supercomputer that consists of 48 highly similar computers. While they are often referred to as massively parallel, a job that begins on $n = 48$ components in a cluster will finish only if all $n$ components function correctly for the duration of the computational task. Essentially, these 48 computers behave as 48 repairable systems in series. Figure 4 plots the cumulative number of failures versus time for each of the 48 computers. There is one "outlying" computer with considerably more failures. In particular, computer 21 is different in both structure and usage.

Whereas this is a repairable system, we seek to establish a stochastic point process, $N(a, b)$, for the number of failures in an interval $(a, b]$. We further define $N(t)$ as the number of failures in $(0, t]$. An important class of models for failure times of a repairable system is that of the nonhomogeneous Poisson processes (NHPP). An NHPP is defined by its nonnegative intensity $\nu(t)$. Under a NHPP:

- The process $N(a, b)$ is a Poisson random variable with mean $\mu(a, b) = \int_a^b \nu(t)\, dt$.
- The processes $N(a_1, b_1)$ and $N(a_2, b_2)$ are independent if $(a_1, b_1)$ and $(a_2, b_2)$ are disjoint (i.e., either $b_1 < a_2$ or $b_2 < a_1$).

Power law process (PLP) and loglinear process models are common choices for the intensity function $\nu(t)$. Ryan and Reese (2005) introduced an extended model that includes a positive parameter $\rho$ to model appropriate asymptotic behavior. They considered intensities of the form

$$\nu(t) = \frac{\phi}{\eta}\left(\frac{t}{\eta}\right)^{\phi-1} + \rho.$$

When $\phi < 1$, the system undergoes reliability growth and has a limiting failure rate of $\rho$. (The intensity never increases or levels off to a constant value, regardless of the choice of $\phi$.)

We present a hierarchical Bayesian model for the Poisson process that governs these data. Assume that the number of failures experienced by computer $i$ in month $j$, $X = [x_{ij}]$ (a $C \times M$ matrix), has probability mass function

$$p(X|\underline{\eta}, \underline{\phi}, \underline{\rho})$$
$$= \prod_{i=1}^{C}\left[\prod_{j=1}^{M}\left[\left\{\left(\frac{tj}{\eta_i}\right)^{\phi_i} - \left(\frac{t(j-1)}{\eta_i}\right)^{\phi_i}\right.\right.\right.$$
$$\left.\left. + t\rho_i\right\}^{x_{ij}}\Big/ x_{ij}!\right]$$
$$\left.\cdot \exp\left\{-\left(\left(\frac{Mt}{\eta_i}\right)^{\phi_i} + Mt\rho_i\right)\right\}\right].$$

Next, allow for a gamma prior distribution on $\underline{\eta}$ that is parameterized in terms of the mean $\mu_\eta$ and standard deviation $\sigma_\eta$. That is, use

$$p(\underline{\eta}|\mu_\eta, \sigma_\eta) = \left(\frac{(\mu_\eta/\sigma_\eta^2)^{(\mu_\eta/\sigma_\eta)^2}}{\Gamma((\mu_\eta/\sigma_\eta)^2)}\right)^C$$
$$\cdot \prod_{i=1}^{C}\left[\eta_i^{(\mu_\eta/\sigma_\eta)^2-1}\exp\left(-\frac{\mu_\eta}{\sigma_\eta^2}\eta_i\right)\right].$$

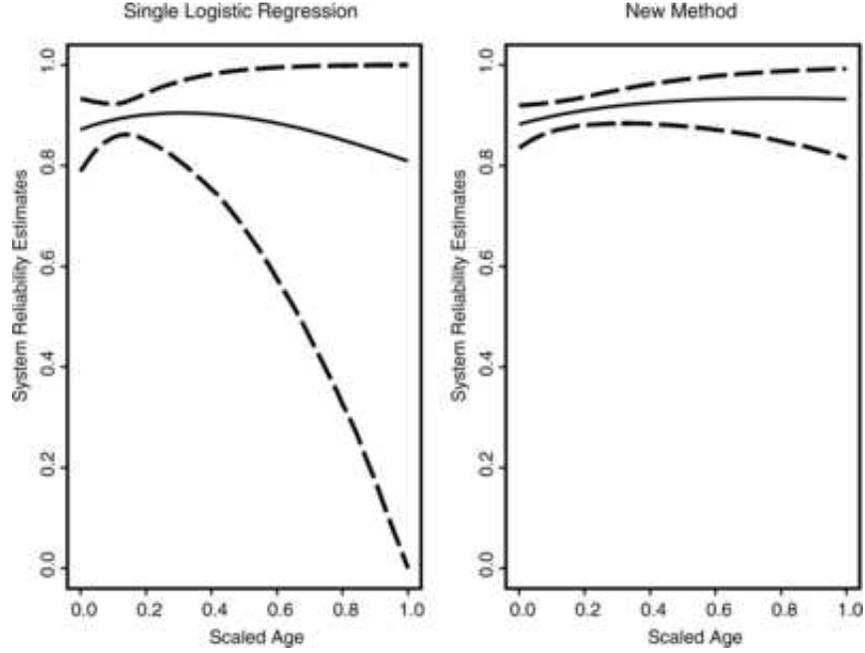

FIG. 3. *Comparison of two reliability estimation procedures. Left: Logistic regression on full system data alone. Right: Integration of component specification tests and partially informative system tests. Shown are posterior medians and 5th and 95th percentiles of posterior distributions.*

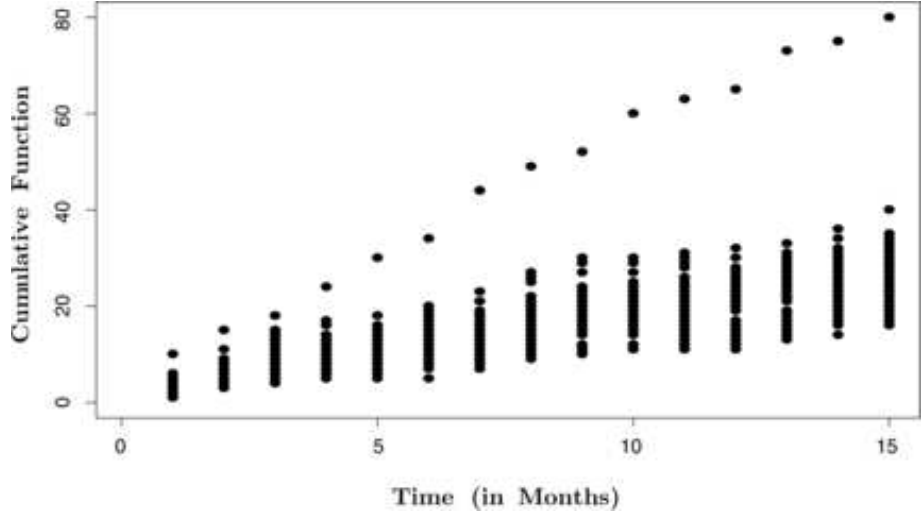

FIG. 4. *Empirical cumulative failure counts of 48 components.*

Similarly, let

$$p(\underline{\phi}|\mu_\phi, \sigma_\phi) = \left(\frac{(\mu_\phi/\sigma_\phi^2)^{(\mu_\phi/\sigma_\phi)^2}}{\Gamma((\mu_\phi/\sigma_\phi)^2)}\right)^C$$
$$\cdot \prod_{i=1}^{C}\left[\phi_i^{(\mu_\phi/\sigma_\phi)^2-1}\exp\left(-\frac{\mu_\phi}{\sigma_\phi^2}\phi_i\right)\right]$$

and

$$p(\underline{\rho}|\mu_\rho, \sigma_\rho) = \left(\frac{(\mu_\rho/\sigma_\rho^2)^{(\mu_\rho/\sigma_\rho)^2}}{\Gamma((\mu_\rho/\sigma_\rho)^2)}\right)^C$$
$$\cdot \prod_{i=1}^{C}\left[\rho_i^{(\mu_\rho/\sigma_\rho)^2-1}\exp\left(-\frac{\mu_\rho}{\sigma_\rho^2}\rho_i\right)\right].$$

This hierarchical specification assumes a priori conditional independence of the computer-specific pa-rameters. This assumption is not as restrictive as that of complete independence. In fact, a posteriori, the parameters will reflect dependence as manifest by the data. As such, we are willing to make this assumption.

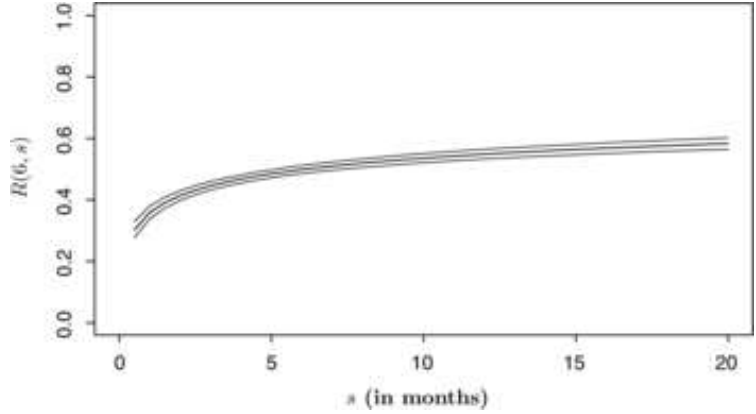

Fig. 5. *Posterior distributions of six-hour reliability versus start time. Shown are the posterior median and 0.05 and 0.95 posterior quantiles.*

The distribution of $(\underline{\eta}, \underline{\phi}, \underline{\rho} | \mu_\eta, \sigma_\eta, \mu_\phi, \sigma_\phi, \mu_\rho, \sigma_\rho)$ has density

$$p(\underline{\eta}, \underline{\phi}, \underline{\rho} | \mu_\eta, \sigma_\eta, \mu_\phi, \sigma_\phi, \mu_\rho, \sigma_\rho)$$
$$= p(\underline{\eta} | \mu_\eta, \sigma_\eta) p(\underline{\phi} | \mu_\phi, \sigma_\phi) p(\underline{\rho} | \mu_\rho, \sigma_\rho).$$

Finally, let

$$\mu_\eta \sim \text{Weibull}(a_{\mu_\eta}, b_{\mu_\eta}),$$
$$\sigma_\eta \sim \text{Weibull}(a_{\sigma_\eta}, b_{\sigma_\eta}),$$
$$\mu_\phi \sim \text{Weibull}(a_{\mu_\phi}, b_{\mu_\phi}),$$
$$\sigma_\phi \sim \text{Weibull}(a_{\sigma_\phi}, b_{\sigma_\phi}),$$
$$\mu_\rho \sim \text{Weibull}(a_{\mu_\rho}, b_{\mu_\rho}),$$
$$\sigma_\rho \sim \text{Weibull}(a_{\sigma_\rho}, b_{\sigma_\rho}).$$

For the $i$th computer, $N_i(a, b)$ is a Poisson random variable with mean $\mu_i(a, b)$. Thus, the probability that the $i$th computer has no failures in $(a, b)$ is

$$P(N_i(a, b) = 0 | \phi_i, \eta_i, \rho_i)$$
$$= 1 - \exp\left(\left(\frac{a}{\eta_i}\right)^{\phi_i} + \rho_i a - \left(\frac{b}{\eta_i}\right)^{\phi_i} - \rho_i b\right).$$

We use an operational definition of reliability to mean *a job of length $l$ run on a computer of age $s$ finishes without computer failure.* Since the supercomputer is a series system in its 48 components, reliability $R(l, s | \Theta)$ is

$$R(l, s | \Theta) = \prod_{i=1}^{48} \left[ \exp\left(\left(\frac{s}{\eta_i}\right)^{\phi_i} - \left(\frac{s+l}{\eta_i}\right)^{\phi_i} - \rho_i l\right)\right].$$

Figure 5 summarizes the posterior distribution of reliability $R(6, s | \Theta)$ versus start time $s$ for six-hour computer runs. The three lines included on this plot are the 0.05 and 0.95 quantiles and the median of $R(6, s | \Theta)$ with respect to $\pi(\Theta | D)$. As $s$ increases, these three lines increase, indicating reliability growth.

While this is a simple system example, it illustrates the power of Bayesian hierarchical models for integrating multiple, similar sources of information to assess overall system reliability. The next example considers a simple system composed of very different components and the combination of system and component testing.

### 3.4 Lifetime Data

As a demonstration of the multiple levels of data collected on a simple system, consider a system that consists of only three components that are all required to work so that the system as a whole works. An event tree representation of such a system is pictured in Figure 6. While this is a simple system, important data combination methods can be illustrated. There are four reliability functions of interest: one for each of the three components and one additional reliability function, which is the system reliability function. Furthermore, suppose that at each component we conduct $n_i = 20, i = 2, 3, 4$, tests and record the time until failure. We also collect $n_S = 10$ full system tests independent of the component data and observe the time until failure. Given this system structure and the test data, we can explore the features of the proposed Bayesian system reliability modeling.

Goodness-of-fit techniques revealed that a reasonable model for the distribution of failure times of the components is Weibull, that is,

$$f_i(t|\Theta) = \frac{\alpha_i}{\beta_i}\left(\frac{t}{\beta_i}\right)^{\alpha_i}\exp\left[-\left(\frac{t}{\beta_i}\right)^{\alpha_i}\right], \quad i = 2, 3, 4.$$

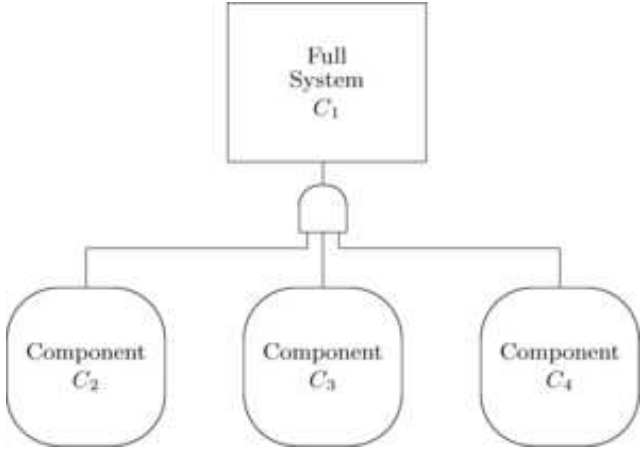

FIG. 6.  *Reliability event tree for system reliability.*

Note that this parameterization of the Weibull distribution is different than that in Section 3.1, with

$\lambda_0 = (1/\beta_i)_i^\alpha$ and $\lambda_1 = \alpha_i$. Here the component reliabilities $R_i(t|\Theta)$ $(i = 2, 3, 4)$ are given by $\int_t^\infty f(t|\Theta)\,dt$, so that the system reliability at time $t$ is $R_S(t|\Theta) = R_2(t|\Theta)R_3(t|\Theta)R_4(t|\Theta)$. Our prior specification is that $\alpha_i$ and $\beta_i$ are all exchangeable (i.e., independent given their prior parameters) and are from a common gamma distribution, that is,

$$p(\alpha_i|\lambda_a, \zeta_a) \propto \alpha_i^{\lambda_a - 1}\exp(-\zeta_a\alpha_i),$$
$$p(\beta_i|\lambda_b, \zeta_b) \propto \alpha_i^{\lambda_b - 1}\exp(-\zeta_b\alpha_i).$$

Then, to complete the hierarchical specification, we propose that $\lambda_a, \zeta_a, \lambda_b$ and $\zeta_b$ have exponential distributions, each with their own rate parameters.

Given the specification above, we use a successive substitution MCMC procedure where each component of the joint posterior distribution is updated one at a time. The posterior distributions (as a function of time) for the reliability function of each of the components in the example system are presented in Figure 7. They are organized as upper left, the posterior distribution of the full system $C_1$; upper right, the posterior distribution for component $C_2$; lower left, the posterior distribution for component $C_3$; lower right, the posterior distribution for component $C_4$.

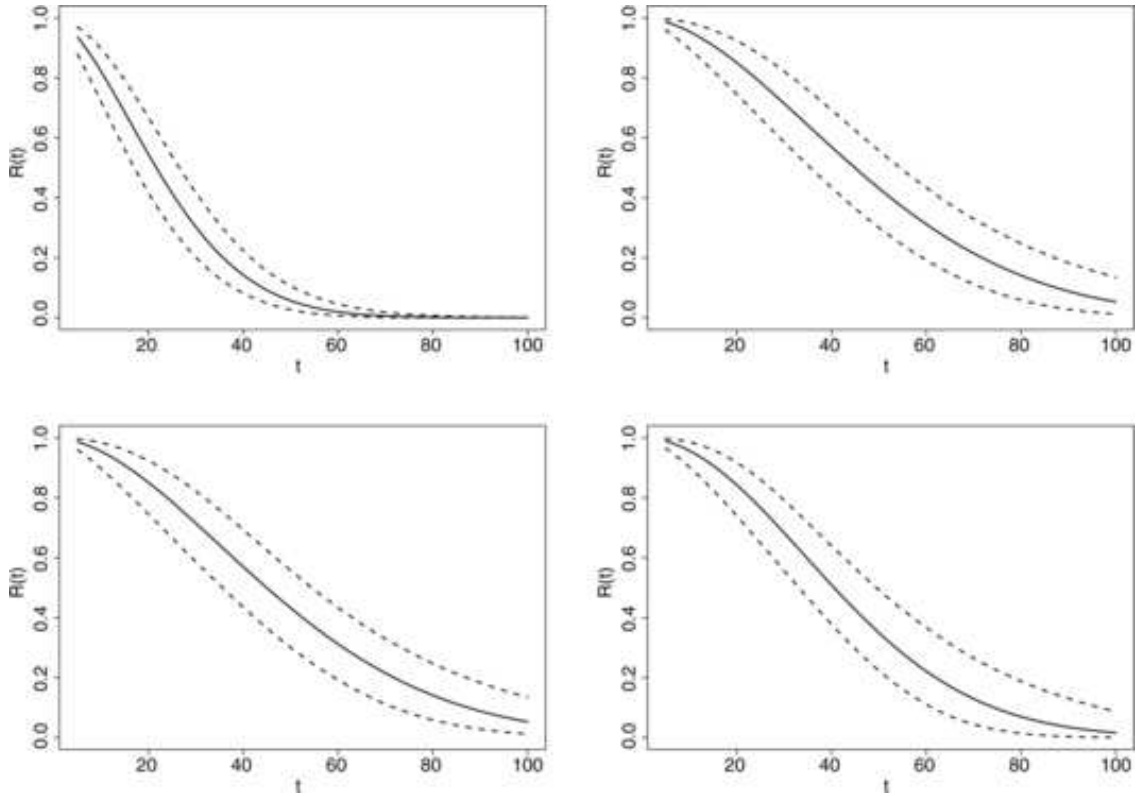

FIG. 7.  *Posterior distributions (as a function of time) for the reliability function of each of the components in the system. The upper left panel is the posterior distribution of the full system $C_1$; the upper right panel is the posterior distribution for the component $C_2$; the lower left panel is the posterior distribution for component $C_3$; and the lower right panel is the posterior distribution for component $C_4$.*

We note, in particular, that the posterior distribution on the system reliability function is less variable than that of any of the components. We only used ten system tests and 20 component level tests, suggesting that the component testing has improved our state of knowledge about the system. Further, we note that the improvement does not reflect an improvement of the magnitude expected if we add 60 (or even 20) full system tests. This would result in a posterior distribution with much less uncertainty. Therefore, the component testing does not inform the posterior proportionately to a full system test, but it does improve our knowledge and can be particularly helpful when full system tests are sparse.

### 3.5 Elicitation for Reliability

The issues around elicitation for reliability fall into three categories: elicitation methodology and techniques, elicitation for parameters and prior specification in reliability models, and elicitation of system structure and dependencies. The first two are relatively well studied; the last is an open research area.

Kadane and Wolfson (1998) stated that "the goal of elicitation, as we see it, is to make it as easy as possible for subject-matter experts to tell us what they believe, in probabilistic terms, while reducing how much they need to know about probability theory to do so." There is emerging consensus that the following assertions represent good technique for parameters and prior elicitation (Kadane and Wolfson, 1998, page 4):

1. Experts should be asked to assess only observable quantities, conditioning only on covariates (which are also observable) or other observable quantities.

2. Experts should not be asked to estimate moments of a distribution (except possibly the first moment); they should be asked to assess quantiles or probabilities of the predictive distribution.

3. Frequent feedback should be given to the expert during the elicitation process. The feedback can be graphical or verbal, and it should help the expert develop coherent probabilities and understand the implications of previous choices.

4. Experts should be asked to give assessments both unconditionally and conditionally on hypothetical observed data.

The psychological underpinnings of these recommendations are summarized in Meyer and Booker (2001). Specific statistical techniques for deriving predictive distributions that are useful in reliability and calculating parameter and hyperparameter distributions from elicited quantities can be found in Kadane and Wolfson (1998), Percy (2002) and Gutiérrez-Pulido, Aguirre-Torres and Christen (2005). More detailed elicitation case studies appear in Keeney and von Winterfeldt (1991) and O'Hagan (1998).

Elicitation of priors for component parameters for systems reliability is especially difficult because, given a fault tree or reliability block diagram structure, the prior distributions for parameters at the components induce prior distributions on the system. For example, suppose that we have a series system with component reliability $p_i$ and that we assume a Uniform$(0, 1)$ prior for each $p_i$. This does not imply that there is a uniform prior on the system itself. Given a series system with $k$ components, the prior distribution on the system is $[\Gamma(k)]^{-1}(-\log p)^{k-1}$, which has mean $2^{-k}$ (Parker, 1972). If the system reliability has a Uniform$(0, 1)$ distribution and we assume that each of the $k$ components has the same prior distribution, then this prior distribution is $[\Gamma(1/k)]^{-1}(-\log p)^{-(k-1)/k}$, which has mean $2^{-1/k}$.

The elicitation of system structure and dependencies among components and failure modes is an emerging area of research. Neil, Fenton and Nielson (2000), Lee (2001) and Wilson, McNamara and Wilson (2007) discussed the construction of Bayesian network representations (Section 4) for complex systems. Seshasai and Gupta (2004) discussed the modeling of structure and information within engineering design process. Klamann and Koehler (2005) proposed qualitative methods for the determination of system structure. The issues are the determination of the correct granularity for representing components and functionality, and the appropriate dependencies among the components, functions and failure modes. Qualitative models of systems that capture these features underlie the development of quantitative statistical models for systems reliability.

## 4. REPRESENTING SYSTEMS

Fault trees and reliability block diagrams are the most common representations in system reliability analysis. However, there are situations where these models do not offer enough flexibility to capture features of the system. Bayesian networks generalize fault trees and reliability block diagrams by allowing components and subsystems to be related by conditional probabilities instead of deterministic AND and OR relationships. Flowgraph models are multistate models that simplify the analysis of time-to-event data.

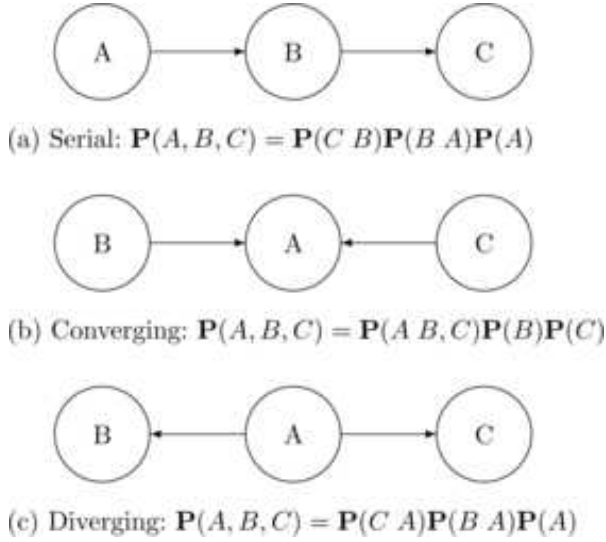

Fig. 8. *Specifying joint probability distributions using a Bayesian network.*

## 4.1 Bayesian Networks

There is growing literature on the use of Bayesian networks (BNs) in reliability (e.g., Portinale, Bobbio and Montani, 2005; Sigurdsson, Walls and Quigley, 2001; Lee, 2001), although there is quite a broad literature on using BNs for probabilistic modeling (e.g., Spiegelhalter, 1998; Neil, Fenton and Nielson, 2000; Laskey and Mahoney, 2000; Jensen, 2001).

Formally, a BN is a pair $N = \langle (V, E), P \rangle$, where $(V, E)$ are the nodes and edges of a directed acyclic graph and $P$ is a probability distribution on $V$. Each node contains a random variable, and the directed edges between them define conditional dependences/independences among the random variables. Figure 8 summarizes the three probabilistic relationships that can be specified in a BN. The key feature of a BN is that it specifies the joint distribution $P$ over the set of nodes $V$ in terms of conditional distributions. In particular, the joint distribution of $V$ is given by

$$\prod_{v \in V} \mathbf{P}(v \mid \text{parents}[v]),$$

where the parents of a node are the set of nodes with an edge pointing to the node. For example, in the serial structure in Figure 8(a), the parent of node $C$ is node $B$, and node $A$ has no parents.

Bayesian networks can be used as a direct generalization of fault trees. The fault tree translation to a BN is straightforward, with the basic events that contribute to an intermediate event represented as parents and a child. Figure 9 shows the correspondence between a fault tree AND gate and a BN converging structure. Notice that a fault tree implies specific conditional probabilities. The same BN converging structure works for an OR gate, with the appropriate conditional probabilities.

Suppose that we have the BN from Figure 10 and suppose that we are interested in calculating the posterior probability for each component and the full system. Hamada et al. (2004) discussed how to approach this problem for the special case of fault trees.

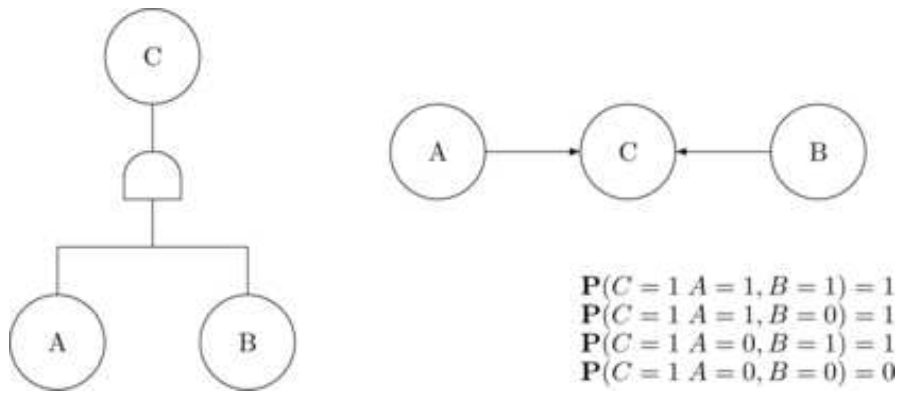

Fig. 9. *Fault tree conversion to a Bayesian network.*

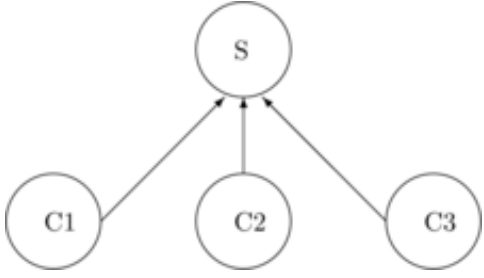

Fig. 10. *Bayesian network generalization of the system example.*

Suppose that we have the same data as given in Section 3.1. However, instead of a series system, we have the relationships

$$\mathbf{P}(S = 1 | C1 = 1, C2 = 1, C3 = 1) = 0.9,$$
$$\mathbf{P}(S = 1 | C1 = 0, C2 = 1, C3 = 1) = 0.4,$$
$$\mathbf{P}(S = 1 | C1 = 1, C2 = 0, C3 = 1) = 0.3,$$
$$\mathbf{P}(S = 1 | C1 = 1, C2 = 1, C3 = 0) = 0.5,$$
$$\mathbf{P}(S = 1 | C1 = 0, C2 = 0, C3 = 1) = 0.1,$$
$$\mathbf{P}(S = 1 | C1 = 1, C2 = 0, C3 = 0) = 0.05,$$
$$\mathbf{P}(S = 1 | C1 = 0, C2 = 1, C3 = 0) = 0.25,$$
$$\mathbf{P}(S = 1 | C1 = 0, C2 = 0, C3 = 0) = 0.$$

Again, drawing from Section 3.1, let $p_1(t) = \text{logit}^{-1}(\theta_0 + \theta_1 t)$, $p_2(t) = \exp(-\lambda_0 t^{\lambda_1})$ and $p_3(t) = \{1 - \Phi(\{\log t - \mu - \log(\alpha - D)\}/\sigma_b)\}$. Then the system data can be modeled as Binomial$(p_S(t))$, where

$$
\begin{aligned}
p_S(t) = {}& 0.9 p_1(t) p_2(t) p_3(t) \\
& + 0.4(1 - p_1(t)) p_2(t) p_3(t) \\
& + 0.3 p_1(t)(1 - p_2(t)) p_3(t) \\
& + 0.5 p_1(t) p_2(t)(1 - p_3(t)) \\
& + 0.1(1 - p_1(t))(1 - p_2(t)) p_3(t) \\
& + 0.05 p_1(t)(1 - p_2(t))(1 - p_3(t)) \\
& + 0.25(1 - p_1(t)) p_2(t)(1 - p_3(t)).
\end{aligned}
$$

Here we omit the dependence on $\Theta$ for space reasons. Figure 11 shows the posterior distribution for system reliability: the posterior mean is solid and the 5th and 95th percentiles are dotted.

The example given above can easily be generalized to the situation where the conditional probabilities are not known, but are described by a distribution. Neil, Fenton and Nielson (2000) and Wilson, McNamara and Wilson (2007) discussed the construction of system models for BNs in detail. For additional examples, see Farrow, Goldstein and

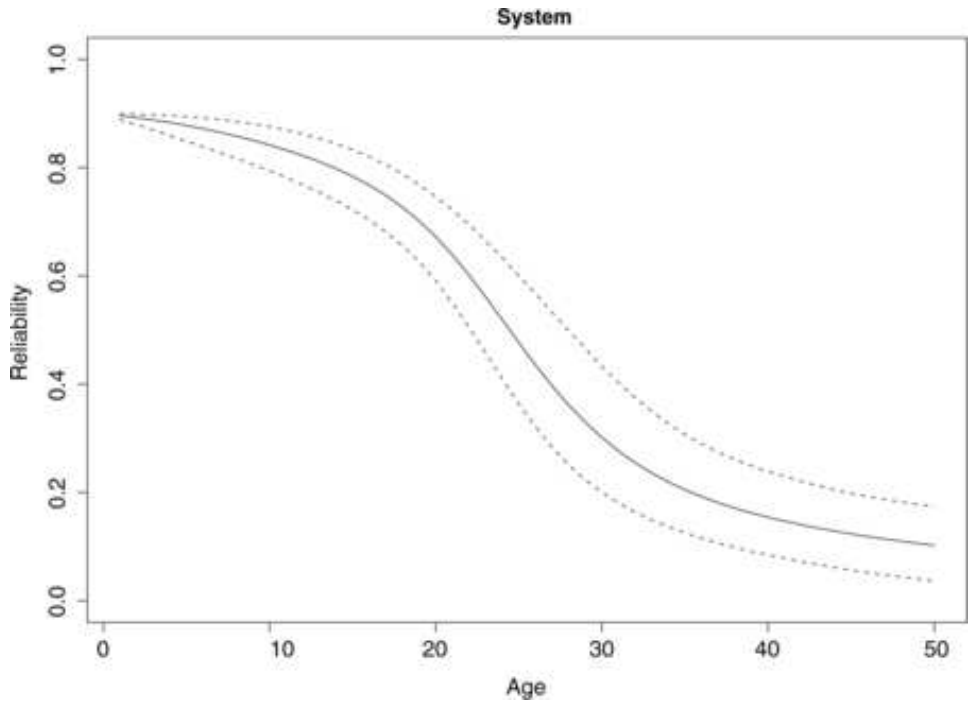

Fig. 11. *Reliability estimates with uncertainty bands for the Bayesian network example. The solid curve is the posterior mean and the dotted curves are the 5th and 95th percentiles of the posterior distribution.*

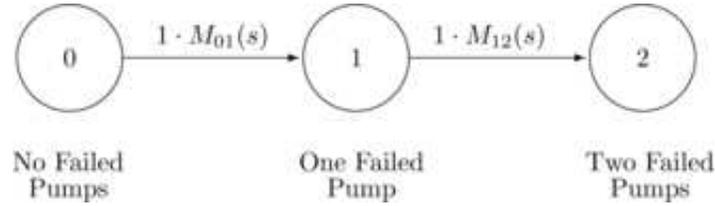

Fig. 12. *Flowgraph model for a series pump system.*

Spiropoulos ([1997](#)), Bedford and Cooke ([2001](#)) and Portinale, Bobbio and Montani (2005).

### 4.2 Flowgraph Models

Flowgraph models offer another representation that can be useful in solving reliability problems. A flowgraph model is one type of multistate model. It is useful for capturing potential outcomes probabilities of outcomes, and waiting times for outcomes to occur, and is often used to model time-to-event data. Like a graphical model, a flowgraph consists of nodes and arcs. However, in a flowgraph model, the nodes (or states) represent outcomes. This differs from a BN, where the nodes represent random variables.

Consider Figure [12](#) from Huzurbazar ([2005](#)). This flowgraph models the states of a pump system with two pumps. The pumps operate independently and the system can operate with one pump if necessary. The nodes of the system represent states of the system: state 0 represents no failed pumps, state 1 represents one failed pump and state 2 represents two failed pumps. One quantity of interest is the time to total failure, or the total time to transition from state 0 to state 1 to state 2.

The directed line segments in a flowgraph are *branches*. Each branch has a transition probability and waiting time distribution associated with a transition from its beginning to ending nodes. The branches are labeled with *transmittances*, each of which is the transition probability multiplied by the moment generating function of the waiting time distribution. For example, in Figure [12](#), the transition probability from state 0 to state 1 is 1.0 and the moment generating function of the waiting time distribution is $M_{01}(s)$. In Figure [13](#), the transition probability from state 1 to state 0 is $p_{10}$ and from state 1 to state 2 is $p_{12}$, where $p_{10} + p_{12} = 1$. In this example, there is no probability of staying in state 1—eventually a transition always occurs.

Suppose that in Figure [12](#) the pumps fail independently with an exponential distribution with mean $1/\lambda_0$, Exponential($\lambda_0$). The transition from state 0 to state 1 happens when either of the pumps fails, which means that its waiting time is the minimum of two independent exponential distributions, which has an Exponential($2\lambda_0$) distribution.

Once in state 1, we assume that the remaining pump has a failure time with an Exponential($\lambda_1$) distribution, with $\lambda_1 > \lambda_0$. We assume that $\lambda_1 > \lambda_0$ to account for the extra stress on the pump once the first has failed. The waiting time to transition from state 0 to state 1 to state 2 is the sum of independent exponential distributions. Since the waiting times are independent, the moment generating function of the sum of the waiting times is the product of the individual times. The moment generating function for an Exponential($\lambda$) is $M(s) = \lambda/(\lambda - s)$ for $s < \lambda$. The moment generating function for the transition from state 0 to state 2 is

$$M_{02}(s) = M_{01}(s)M_{12}(s)$$

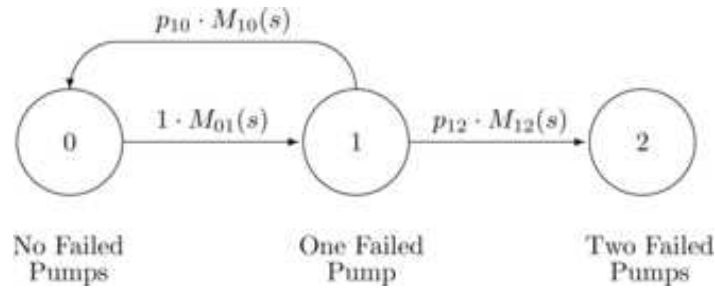

Fig. 13. *Flowgraph model for a series pump system with feedback.*

$$= \frac{2\lambda_0}{2\lambda_0 - s} \frac{\lambda_1}{\lambda_1 - s}.$$

This moment generating function uniquely determines the distribution of the waiting time. Since we can now write an equivalent flowgraph with the transmittance from state 0 to state 2, we have *solved* the flowgraph from 0 to 2. Huzurbazar (2005) gave a general algorithm based on Mason's rule to solve flowgraphs like those in Figures 12 and 13. The moment generating functions in the transmittances can be converted to probability density functions (or other summaries, like reliability or hazard functions) either analytically or using saddlepoint approximation techniques. Huzurbazar (2000, 2005) gave a number of examples of flowgraph models, solving flowgraphs and inverting flowgraph moment generating functions.

## 5. RESOURCE ALLOCATION

Sections 2 and 3 considered the analysis of various sources of component data and a mix of component and system data, respectively, to assess system reliability. In this section, we address how to allocate limited testing resources; we simply refer to this problem as resource allocation (Hamada et al., 2004). That is, given a limited budget, where should additional tests be done (at the system level and/or the component level) and how many tests should be done there?

First, we assume that there is a cost for collecting additional data and that it is more costly to collect higher level (e.g., system) data than lower level (e.g., component) data. For specified costs, a candidate allocation, that is, the number of tests at the system level and at all the components, must not exceed a fixed budget.

Next, we need a criterion with which to evaluate a candidate allocation and to compare two different candidate allocations. We use one based on repeated pre-posterior analyses. The fact that we can analyze the varied data presented in Sections 2 and 3 allows us to take such an approach. The criterion can be described operationally as follows. We draw from the parameter prior distribution (based on the existing data), simulate data according to the candidate allocation using the current prior draw as the true parameters and then update with the simulated data to obtain the parameter posterior distribution. Using this parameter posterior distribution, we evaluate the system reliability posterior distribution and record some distributional characteristic. For example, we often use the length of the central 90% credible interval as a measure of uncertainty that we would like to reduce. Repeating this procedure produces an empirical distribution of posterior credible interval lengths. As a criterion for the candidate allocation, we use an upper quantile, for example, the 0.90 quantile.

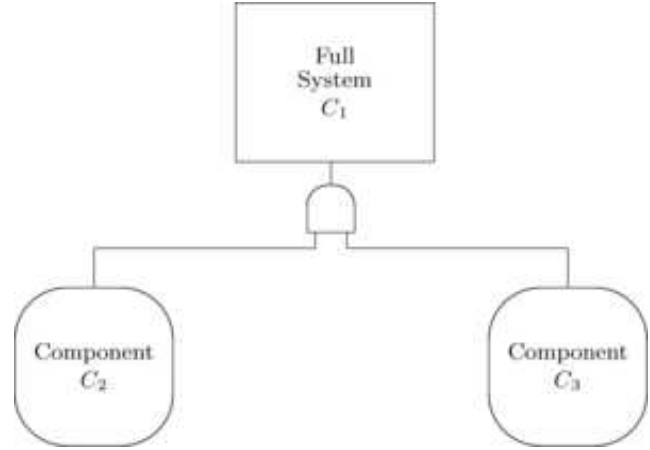

FIG. 14. *Event tree for a two-component series system.*

Finally, we need to find the candidate allocation that optimizes, in this case minimizes, the criterion. To do the optimization problem, we can use a genetic algorithm (GA) (Goldberg, 1989). We have implemented a GA for resource allocation in R (R Development Core Team, 2004), which generates the candidate allocations. A candidate allocation is also evaluated in R by repeatedly generating data sets and calling YADAS (Graves, 2003, 2007) to do the Bayesian updating. YADAS produces an output file of parameter posterior draws that is read into R to calculate the candidate allocation criterion.

In the remainder of this section, we consider the case where there are only binomial count data at the system and component levels. We illustrate resource allocation for a simple series system that consists of two components as displayed in Figure 14 as an event tree.

Johnson, Graves, Hamada and Reese (2003) showed how to combine system and component level binomial data in a reliability assessment. For example, if the series system structure in Figure 14 is valid and the components are independent, then the system reliability $p_1$ equals $p_2 p_3$, the product of the two component reliabilities. Consequently, system level data are informative about the component reliabilities through this relationship.

Let $TC_i$ be the corresponding costs: $TC_1$ is the cost of a system test, and $TC_2$ and $TC_3$ are the component costs. Let $n_i$ be the corresponding number of tests so that for budget $B$, $\sum_{i=1}^{3} TC_i n_i \leq B$.

Under this scenario where the system structure (i.e., series) holds and binomial count data are collected, resource allocation depends on these costs. If $TC_1 \geq TC_2 + TC_3$, then the optimal allocation will consist only of component tests. Even if $TC_1 = TC_2 + TC_3$, there is still more information gained from individual component tests than one system test. That is, doing one system test as compared with testing each component once provides less information. If $TC_1 = TC_2 = TC_3$, then the optimal allocation is all system tests. If $TC_1 < TC_2 + TC_3$, there must be a mixture of system and component tests, but trying to characterize this mixture remains to be done.

An important reason for performing system tests is that they are integrative, which in the above discussion was not accounted for. That is, does the system work when all the components are assembled? Consequently, system tests are needed to assess the assumed structure. The previously stated relationship between the system and component reliabilities for the simple series system,

$$p_1 = p_2 p_3,$$

assumes that the series structure with independence holds. To allow for the possibility that the assumed structure does not hold perfectly, consider the relationship

$$(3) \qquad p_1 = p_2 p_3 / [p_2 p_3 + (1 - p_2 p_3) \exp(-\beta)].$$

Here $\beta$ is a bias term for which $\beta = 0$ means that the series structure with independent components holds; also, for $\beta < 0$, $p_1 < p_2 p_3$ and for $\beta > 0$, $p_1 > p_2 p_3$. Note that if a specific departure from the assumed structure is of interest, the departure can be accommodated. For example, if there is a possible additional failure mode due to common causes, the relationship given in Mosleh (1991) can be used.

For resource allocation, we see from (3) that to reduce the uncertainty about the system reliability $p_1$, the uncertainty of the bias term $\beta$ also needs to be reduced. To do this requires some system tests.

Consider resource allocation when the assumed system structure may not hold for the following problem:

- The existing data consist of 2 system tests (both successes), 5 component 1 tests (5 successes) and 10 component 2 tests (9 successes).

- Prior distributions on the component reliabilities and bias term $\beta$ are taken to be diffuse. Combined with the existing data listed above, the 90% credible intervals based on the resource allocation prior distributions are (0.83, 1.00) for the component 1 reliability, (0.77, 0.98) for the component 2 reliability and ($-1.56$, 2.75) for the bias term.

- The resulting resource allocation prior distribution for the system reliability has a 90% credible interval of (0.579, 0.992). Consequently, the length of the initial 90% credible interval for system reliability is 0.413.

For a budget $B = 2500$ and costs $TC_1 = 30$ and $TC_2 = TC_3 = 1$, the optimal allocation (based on evaluating 2000 candidate allocations using a GA) is to do as many system tests (i.e., 83) as possible. An allocation of $(n_1, n_2, n_3) = (83, 10, 0)$ yields a value of 0.160 based on 1000 generated data sets with 10,000 posterior draws per data analysis. For this case, we see that in spite of the system test cost being much larger than the component test costs, the entire budget is essentially spent on system tests. If initially there is less uncertainty about the bias term, we expect there to be an allocation between system and components tests; recall that the no bias case presented above allocated the budget entirely to component tests.

More study of resource allocation for more complicated systems is needed. On the other hand, for a specific situation one needs to employ an optimization algorithm such as the GA we used in this example to find an optimal allocation.

## 6. DISCUSSION

In this paper, we hope that we have conveyed the importance of the role that statisticians can play in assessing system reliability today and the many research challenges that it presents. Somewhat facetiously, we thought of titling this paper "This ain't your father's reliability!" or "System reliability assessment—a statistician's playground!," because both express the excitement that we have about the research challenges in this field.

As Sections 2 and 3 showed, novel statistical models arise when statisticians want to leverage information from all available data to bear in an assessment. Even assessing a single component can be challenging when the data are from computer experiments (Santner, Williams and Notz, 2003) in which verification, validation and calibration need to be addressed, and where multiscale physical experiments and historical system tests with multiple measurement errors must be integrated. Moreover, as much engineering and science knowledge as possible needs to be incorporated into the statistical modeling.

Section 3 presented some of the challenges in incorporating multilevel data from various sources in an assessment. Another example occurs when the data come from different tests at different levels, some of which are done at more severe conditions than those experienced in normal use (Reese, Hamada and Robinson, 2005). Section 3.5 also discussed elicitation of expert knowledge. This is critical in capturing both the functional and physical structure of a system, and more research is needed on techniques and tools for carrying out this activity.

In Section 4, richer representations than fault trees and reliability block diagrams were presented. More research is needed on statistical inference with these representations. Section 5 presented the emerging problem of resource allocation. There are many interesting problems beyond the binomial case. For example, in an accelerated degradation data experiment on a single component, one needs to determine how much of the budget should be spent on this experiment and subsequently, what levels of the accelerating variable should be studied, how many units should be tested at each level and how often the units should be inspected. There is much research here that remains to be done.

Implementation of reliability assessment in large systems is an issue and tools are needed, which is a research effort in itself. Our organization (Statistical Sciences Group at Los Alamos National Laboratory) is developing qualitative system representation tools such as GROMIT (Klamann and Koehler, 2005) and statistical modeling tools such as YADAS (Graves, 2003, 2007), as well as an interface between them. However, many challenges remain. For example, system reliability assessments are computationally intensive. What approximations can be incorporated without sacrificing accuracy? Do we need the power of a supercomputer? Resource allocation is even more computationally intensive and brings the issues of computation to the forefront.

## ACKNOWLEDGMENTS

The authors would like to thank Harry Martz, James Lynch and two anonymous reviewers for suggestions that led to substantial improvements of this paper.